# One step growth of GaN/SiO$_2$ core/shell nanowire in vapor-liquid-solid route by chemical vapor deposition technique

B.K. Barick[*], Shivesh Yadav, S. Dhar

*Department of Physics, Indian Institute of Technology, Bombay, Mumbai-400076, India*

**Abstract:** GaN/SiO$_2$ core/shell nanowires are grown by cobalt phthalocyanine catalyst assisted vapor-liquid-solid route, in which Si wafer coated with a mixture of gallium and indium is used as the source for Ga and Si and ammonia is used as the precursor for nitrogen and hydrogen. Gallium in the presence of indium and hydrogen, which results from the dissociation of ammonia, forms Si-Ga-In alloy at the growth temperature ~910 °C. This alloy acts as the source of Si, Ga and In. A detailed study using a variety of characterization tools reveals that these wires, which are several tens of micron long, has a diameter distribution of the core ranging from 20 to 50 nm, while the thickness of the amorphous SiO$_2$ shell layer is about 10 nm. These wires grow along $[10\bar{1}0]$ direction. It has also been observed that the average width of these wires decreases, while their density increases as the gallium proportion in the Ga-In mixture is increased.

*Corresponding author's E-mail addresses: bkbarick@gmail.com

## I. Introduction

GaN nanowires are promising material for high efficiency nanoscale electronic and optoelectronic devices such as light emitting diodes [1], lasers [2] and transistors [3]. As compared to bulk, nanowire geometry has certain advantages. For example, the density of extended defects such as dislocations is found to be much less in these nanostructures than in bulk [4]. Moreover, confinement of carriers in the radial direction in a nanowire can in principle lead to an enhanced carrier mobility due to the decrease of scattering cross-section because of reduced dimensionality. However, these advantages can not be exploited efficiently as the surface plays a very important role in governing their electronic properties. For example, attachment of some adatoms with the surface can adversely affect the carrier mobility as well as influence the luminescence characteristics of nanowires [5]. As a result of this strong surface sensitivity, reliable and reproducible use of devices based on nanowires becomes questionable. One way to overcome this problem is to cover the surface with a shell of a dielectric material. Core/shell nanowires involving different semiconductors such as GaAs/AlGaAs and Si/Ge are found to show improved optical and transport properties as compared to their bare nanowire counterparts. For example, enhancement of radiative emission yield in GaAs/AlGaAs core/shell nanowires as compared to bare GaAs nanowires is attributed to the passivation of the surface defects of the core by the shell layer [6]. Similarly, the suppression of scattering at the surface in core/shell nanowires can lead to the enhancement of the carrier mobility in Ge/Si [7] and GaAs/AlGaAs [8] nanowires. Recently, InGaN/GaN core/shell heterostructure nanowires have received much attention for their potential in LEDs and photo-detection devices [9-11]. Several groups have reported the growth of GaN/AlN core/shell nanowires [12-15]. Growth of these hetrostructures is believed to be the result of two growth processes occurring simultaneously. While, the growth of GaN nanowires takes place through a catalyst-assisted vapor–liquid–solid (VLS) mode, the AlN shell layers are grown on the side walls of the wires through a vapor-solid (VS) growth mode [15].

It should be noted that the semiconductor nanowire based field-effect transistors (NWFET) could be a better alternative to planar metal-oxide-semiconductor field effect transistor (MOSFET) as the former is expected to have higher mobility of carriers due to the reduced dimensionality [16]. Typically, single-nanowire (NW) GaN FETs are bottom gated. These devices are fabricated through dispersing NW on a conducting substrate (typically n- or p- type Si) coated with dielectrics such as $SiO_2$ [17,18], $HfO_2$ [19], $GaO_2$ [20]. While, for the top gated devices, the dielectric layers are deposited on the top of NWs. However, it is difficult to achieve a uniform gate coupling with the wires using above methods. In fact, better gate coupling could be possible by a cylindrical gate dielectric which wraps around the nanowire [21]. But the fabrication of such a surround-gate geometry involves several processing steps and therefore significantly time consuming. A better way will be to cover the NW with a dielectric shell during growth. Gallium oxide shell layers surrounding GaN NWs can be obtained by post growth high-temperature oxidation of the wires [22]. However such a treatment can result in various structural defects in the nanowires leading to the reduction of carrier mobility.

Here we demonstrate a single-step route to grow GaN NWs encapsulated with amorphous $SiO_2$ shell, which can serve as a high-quality gate oxide. The growth is carried out through a Co-phthalocyanine catalyst assisted VLS mechanism. In this process, a Si wafer coated with gallium and indium acts as the source for Ga and Si and ammonia as nitrogen precursor. We believe that in the presence of hydrogen that is generated due to the dissociation of ammonia, silicon dissolves in the mixture of indium and gallium to form Si-Ga-In alloy, which can produce sufficient quantity of Si flux even at 910 °C. A detailed characterization reveals that these wires are several tens of microns long and they grow along $[10\bar{1}0]$ direction. The core of these wires has a diameter distribution ranging from 20 to 50 nm, while the shell thickness is about 10 nm.

## II. Experimental

High purity ammonia gas (99.9995%) was used as precursor for nitrogen. Cobalt phthalocyanine (Co-Ph) (b-form, dye content 97%) was used as the catalyst and argon (99.999%) was used as the carrier gas. 190 mg of indium and gallium mixtures prepared with various weight ratios were used as metal source. A Si wafer coated with In-Ga mixture (wafer-A) was placed inside a cylindrical quartz reactor. 50 mg of cobalt phthalocyanine was dissolved in 1 ml of toluene and casted on another Si substrate (wafer-B), which was then dried under infrared (IR) light and placed next to the wafer-A (downstream side). The quartz tube was then purged with high purity argon (99.999%) at a flow rate of 200 standard cubic centimeters per minute (sccm) for 20 minutes. Subsequently, the temperature of furnace was ramped to a desired temperature of 910 °C at a rate of 34 °C per minute. Meanwhile at temperature 400 °C, 20 sccm of ammonia flow was switched on. Growth was carried out at 910 °C for 5 hours. Several samples were grown with indium to gallium weight ratios of 1:4 (sample A),1:2(sample B), 1:1(sample C), 2:1(sample D) and 4:1(sample E). A reference nanowire sample is grown using only gallium as metal source under the same growth condition (sample F).

Surface morphology of all these samples were analyzed by scanning electron microscopy (SEM) technique. Samples for transmission electron microscopy (TEM) studies were prepared by transferring the nanowires from the substrate to methanol solution, then drop casted the solution on carbon coated Cu TEM grids and dried under IR lamp. Bright field Imaging microscopy, diffraction, high resolution TEM imaging, scanning tunneling electron microscopy (STEM) and energy dispersive X-ray (EDX) spectroscopic studies were performed using a FEG TEM (JEOL JEM 300 kV) operated at an accelerating voltage of 300 kV.

Photoluminescence spectroscopy was carried out at room temperature using a 25 mW He-Cd laser (325 nm) as excitation source. While the emission was analyzed through a 0.5 m focal length monochromator attached with a Peltier cooled CCD array from Andor technology. Raman spectra were recorded in backscattering geometry using the 514.5 nm line of an Ar$^+$ laser covering the range 100–1000 cm$^{-1}$ using laser-Raman spectrometer (Model-RAMNOR HG-2S).

## III. Results and discussions

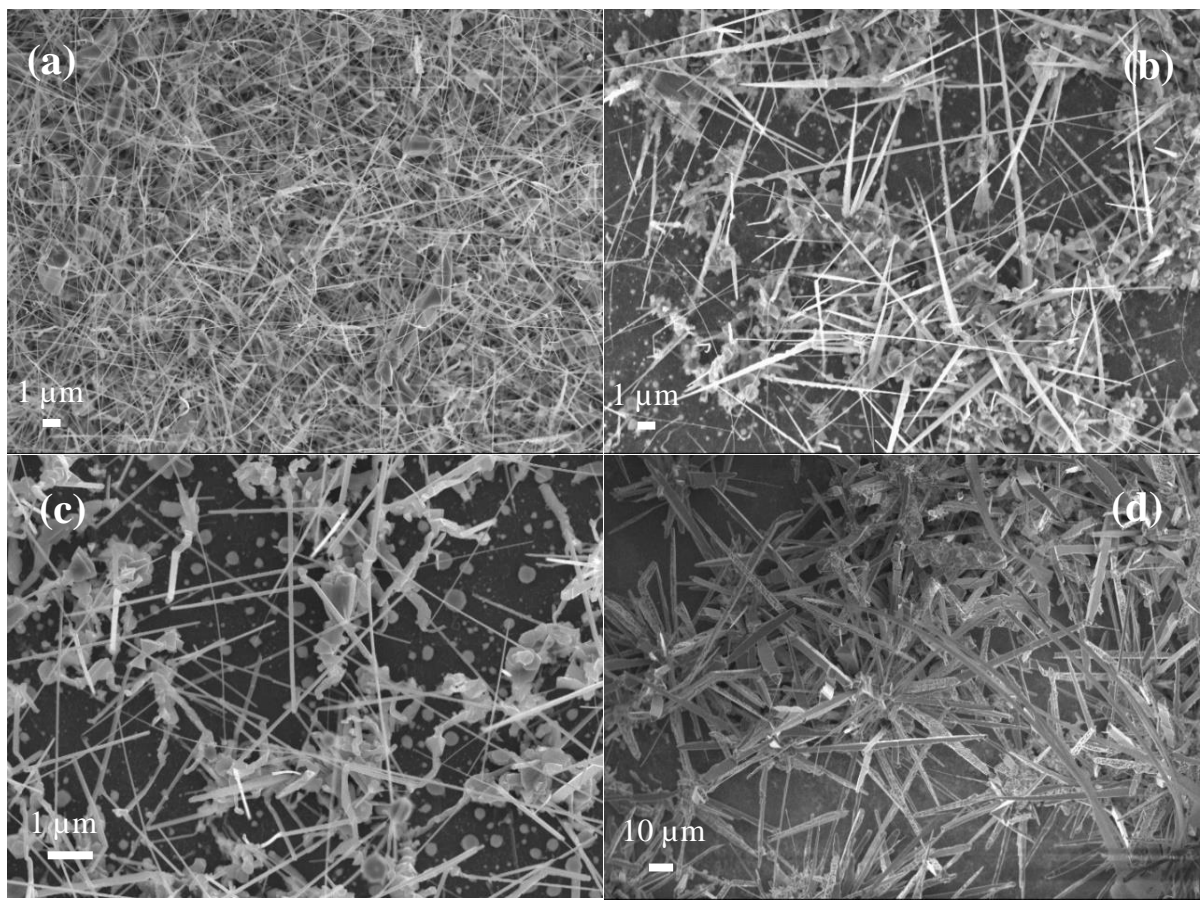

**Figure 1** SEM image of samples grown with In:Ga weight ratio of (a)1:4 (Sample A) (b) 1:2 (Sample B) (c) 1:1 (Sample C) (d) 2:1(Sample D).

Figure 1 shows a top view SEM image of nanowires grown with different indium:gallium weight ratios on Co-phthalocyanine coated Si (100) substrate at 910 °C. Several tens of micron long nanowires with diameter of a few tens of nanometer are grown for the samples A, B, and C as evident from the figure. Noticeably, the density of wires increases with the proportion of Ga in the alloy precursor. It has also been observed that the wires become wider as the Ga proportion in the alloy decreases. For sample D, thick and scattered rod like structures are evident from the figure. When Ga proportion in the alloy is further reduced (sample E), no evidence of nanowires could be found.

Figure 2 (a) shows HRTEM micrographs of GaN nanowires for sample B. HRTEM image shows a core surrounded by a shell part. The width of the nanowire core is found to be about 30 nm and the thickness of shell is about 10 nm. Figure 2 (b) shows a selected area electron diffraction (SAED) pattern for a wire. Diffraction spots show hexagonal symmetry suggesting a wurtzite phase with the zone axis along [0002] direction. Position of these spots, in fact, satisfy wurtzite phase of GaN. Figure 2 (c) represents the bright field high resolution transmission microscopy image for a wire. Core of the wire is crystalline, showing lattice fringes, whereas the shell part is amorphous. In the core part, inter-planar distances, which are determined from the fast furrier transform (FFT) pattern as well as directly from the high resolution micrograph, confirm the wurtzite phase of GaN wire. Furthermore, it has also been found that the growth direction of these wires is $[10\bar{1}0]$, which is non-polar m-plane of wurtzite GaN. It should be noted that core/shell nanowires are found in sample A as well. While, in sample C and D, wires are found to be bare GaN nanowires. Reference sample F is found to be bare GaN nanowires with similar length and diameter distributions as has been observed in sample B. These wires are also found to grow along $[10\bar{1}0]$ direction. However, no shell structure surrounding these wires could be seen [23].

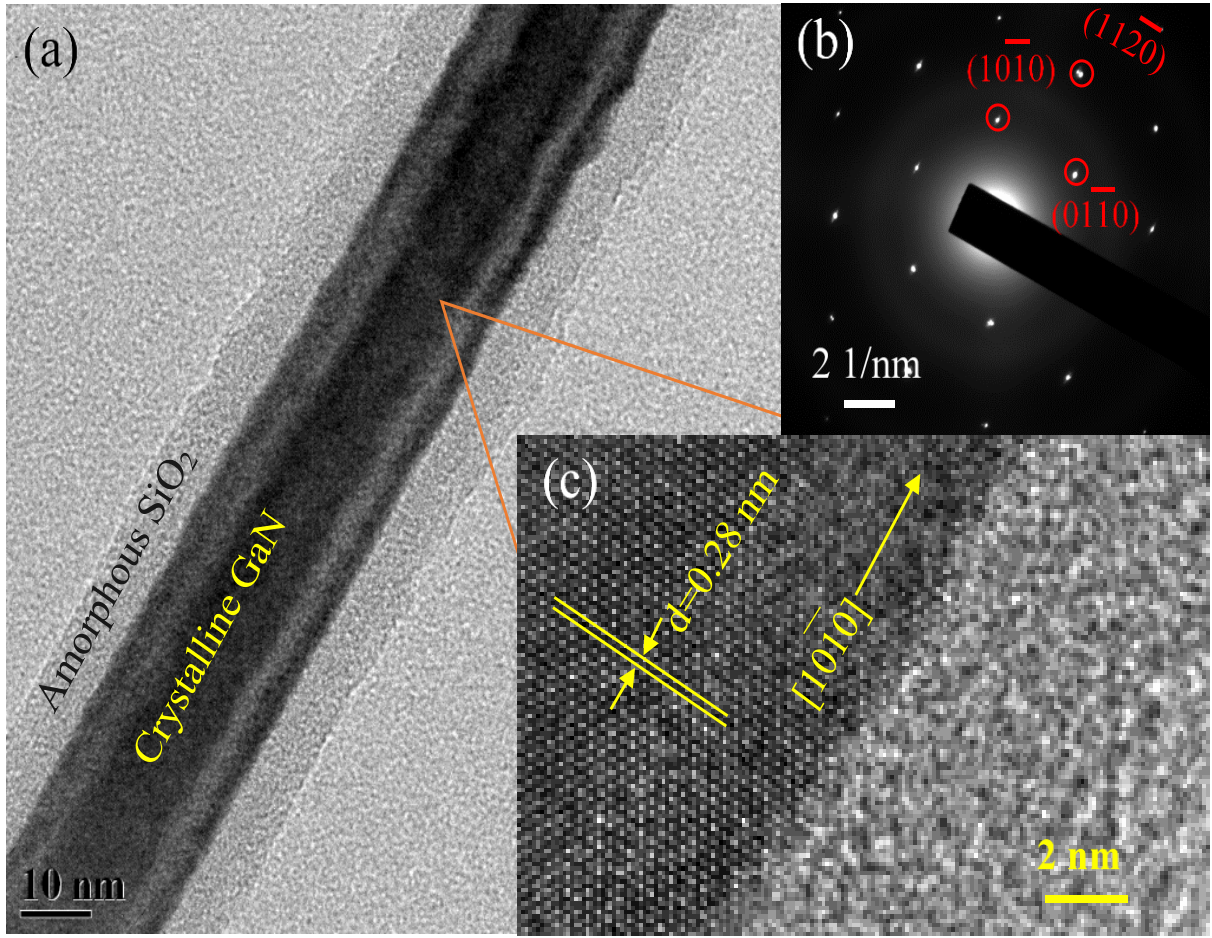

**Figure 2** (a) High-resolution transmission electron micrograph of nanowires, (b) selected area electron diffraction pattern of one of the nanowires and (c) High-resolution TEM image showing lattice planes.

Figure 3(a) compares the room temperature Raman spectra recorded for GaN/SiO$_2$ core/shell nanowires of sample B and for the bare GaN nanowires of the reference sample F. All the characteristic peaks associated with wurtzite phase of GaN, namely the $E_2^{high}$ (568.8 cm$^{-1}$) [24], $A_1^{LO}$ (730.7 cm$^{-1}$) [24], zone boundary (ZB) phonon (255.89 cm$^{-1}$) [25], overtones of acoustic (AO) phonons (323.3 cm$^{-1}$ and 424.5 cm$^{-1}$)[26, 27] along with the surface optical (SO) phonon (691.3 cm$^{-1}$)[28] are visible for both the samples. The feature at 520 cm$^{-1}$ corresponds to the transverse optical (TO) mode of the Si substrate [29]. Figure 3(b) compares the normalized room temperature PL spectra for GaN (sample F) and GaN/SiO$_2$ core/shell nanowires (sample B). Evidently, both the spectra look very similar and featured by a near band edge transition (at about 3.4eV) and a broad but comparatively weak Yellow luminescence (YL) band [at 2 eV]. These observations confirm that the core part of the wire is GaN. In order to identify the shell part of the wire we have carried out an EDX study of individual core/shell nanowires.

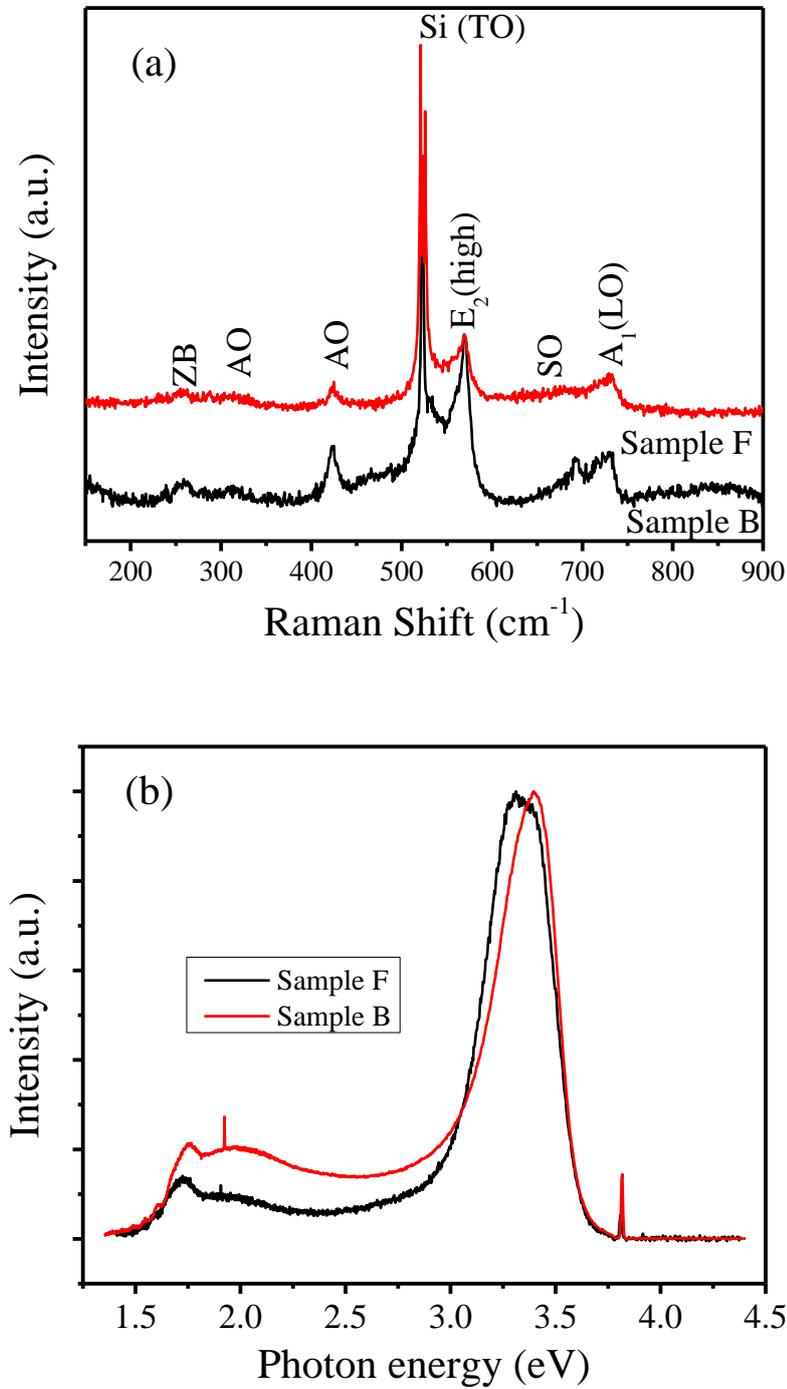

**Figure 3** (a) Raman spectra (b) Photoluminescence spectra of GaN and GaN/SiO$_2$ core/shell nanowire.

Figure 4 (a) shows STEM image of a core/shell nanowire with EDX line scans for different elements. It clearly revels that Ga is present in the core part of the wire while silicon and oxygen are present over the entire wire width, meaning these elements are present in the shell part of the wire. Figure 4 (b) shows EDX area scans recorded on a wire for Si, O, Ga and N. All these results along with the results of Fig. 2 and 3 make us to believe that the core part of the wire is wurtzite GaN, while the amorphous shell part is likely to be $SiO_2$. Interestingly, no trace of indium could be found either in the core or in the shell part of the wire. Question thus arises about the role indium in the formation of these wires.

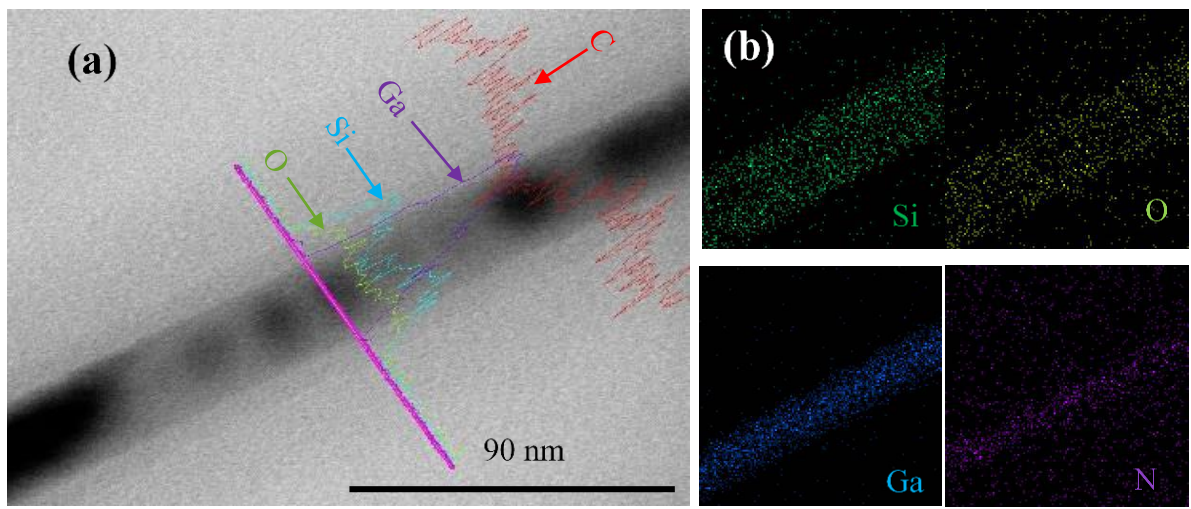

**Figure 4** (a) STEM image of a core/shell nanowire with EDX line scans for different elements. (b) EDX scans recorded on the wire for Si, O, Ga and N.

Note that, we have kept indium and gallium on a Si wafer. In fact, gallium, indium and silicon can form Si-In-Ga alloy at a temperature far less than the melting temperature of Si (1400 °C ) in the presence of hydrogen [30]. Hydrogen, which results from the thermal dissociation of ammonia, is always present during the growth of these wires []. This can help the Si in wafer A to mix with gallium and indium to form Si-In-Ga alloy. We believe, that ternary alloy is thermally evaporated at the growth temperature (910 °C ) and dissolved along with nitrogen in cobalt droplets. Note that at the growth temperature, cobalt phthalocyanine is found to be converted to cobalt droplets, which act as the seeds for the nanowire growth [23]. Once supersaturation is achieved, the (Ga,Si,In)N sublimates in the form of wires following VLS growth mechanism. It is plausible that during the cooling down stage (after the completion of the growth) the (Ga,Si,In)N alloy decomposes in a way that In and Si go to the surface leaving GaN at the core of the wire. Temperature at this stage might be sufficiently high for indium to be fully evaporated from the surface leaving behind a porous shell of Si on the surface as shown schematically in Fig. 4. It is highly likely that when the sample is taken out of the reactor, the surface Si shell layer reacts with atmospheric oxygen to be transformed to $SiO_2$.

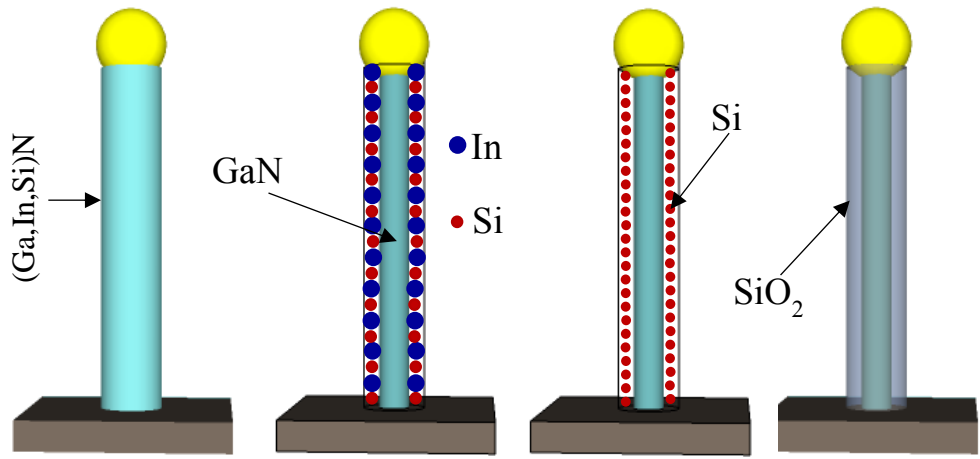

**Figure 4** Schematic representing the formation of GaN/$SiO_2$ core/shell nanowire.

**IV Conclusion**

Our study shows that GaN nanowires encapsulated with amorphous $SiO_2$ shell layers can be grown on Co-phthalocyanine coated clean Si substrates when a Si wafer coated with a mixture of gallium and indium is used as metal source in a VLS growth carried out at a growth temperature of 910 °C under the flow of ammonia. These wires are found to be several tens of microns long and as narrow as several tens of nanometer wide. The GaN core is found to grow along $[10\bar{1}0]$ direction. Average diameter of these wires decreases, while the wire density increases as the gallium proportion of the Ga-In mixture is increased. Nanowires grown with sufficiently high indium to gallium ratios are found to be GaN. However these wires are not found to be encapsulated at all. This method of growing core/shell GaN/$SiO_2$ nanowires could potentially be a viable way to make gated electronic devices based on single nanowires.

**Acknowledgements**

We acknowledge Sophisticated Analytical Instrument Facility (SAIF), IIT Bombay, Center of excellence for nanoelectronics (CEN) for providing various experimental facilities and FIST (Physics)-IRCC central SPM facility of IIT Bombay. This work was supported by Department of Science and Technology (DST) under Grant No: SR/S2/CMP–71/2012 and Council of Scientific & Industrial Research (CSIR) under Grant No: 03(1293)/13/EMR-II, Government of India.